\documentclass[a4paper,11pt,amsmath,amssymb]{article} 
\usepackage{jheppub}
\usepackage{amsmath} 
\usepackage{array,relsize,float}
\usepackage{pstricks}
\usepackage{color} 
\usepackage{amssymb}
\usepackage{slashed}
\usepackage{tikz-cd}
\usepackage{graphicx} 
\usepackage{epsfig}
\usepackage{multicol} 
\usepackage{xspace}
\usepackage[T1]{fontenc}
\usepackage{slashed} 
\usepackage{hyperref}
\usepackage{pdfpages}
\usepackage{array}
\usepackage{tikz-cd}
\usepackage{amsfonts,layout,appendix,subfigure}
\allowdisplaybreaks
\input{epsf} 

\renewcommand{\thefigure}{\arabic{figure}}
 
\def\beq{\begin{equation}} \def\eeq{\end{equation}}
\def\beqn{\begin{eqnarray}} \def\eeqn{\end{eqnarray}}
 
\newcommand\as{\alpha_{\mathrm{S}}} 
\newcommand\aem{\alpha} 
 
\def\beq{\begin{equation}} 
\def\eeq{\end{equation}} 
\def\beqn{\begin{eqnarray}} 
\def\eeqn{\end{eqnarray}} 
 
\def\to{\rightarrow}
\def\nn{\nonumber}

\pdfoutput=1

\newcommand{\valencia}{Instituto de F\'{\i}sica Corpuscular, Universitat de Val\`{e}ncia -- Consejo Superior de Investigaciones Cient\'{\i}ficas, Parc Cient\'{\i}fic, E-46980 Paterna, Valencia, Spain.}
\newcommand{\culiacan}{Facultad  de  Ciencias  F\'isico-Matem\'aticas,  Universidad  Aut\'onoma  de  Sinaloa,  Ciudad  Universitaria, CP 80000 Culiac\'an, M\'exico.}
\newcommand{\berlin}{Deutsches Elektronen-Synchrotron, DESY, Platanenallee 6, D–15738 Zeuthen, Germany.}

\begin{document}

\title{Analysis of the internal structure of hadrons using direct photon production}
\author[a]{David F. Renter\'ia-Estrada,}
\author[a]{Roger J. Hern\'andez-Pinto}
\author[b,c]{and German F. R. Sborlini}
\affiliation[a]{\culiacan}
\affiliation[b]{\berlin}
\affiliation[c]{\valencia}

\emailAdd{davidrenteria.fcfm@uas.edu.mx}
\emailAdd{roger@uas.edu.mx}
\emailAdd{german.sborlini@desy.de}

\preprint{DESY 21-057}

\abstract{Achieving a precise description of the internal structure of hadrons is a hard task, since there are several bottlenecks to obtain theoretical predictions starting from first principles. In order to complement the highly-accurate experiments, it is necessary to use ingenious strategies to impose constraints from the theory side. In this article, we describe how photons can be used to unveil the internal structure of hadrons. Using up-to-date PDFs and FFs, we explore how to describe NLO QCD plus LO QED corrections to hadron plus photon production at colliders.}

\setcounter{page}{1}
\maketitle

\section{Introduction and motivation}
\label{sec:introduction}
Understanding the internal structure of non-fundamental particles implies dealing with complex models, whose solutions can not be easily obtained. Roughly speaking, starting from the accepted framework to describe all the fundamental particles, namely the Standard Model (SM), it is not clear how to model strongly interacting systems from first principles. The typical energy scale associated to these systems is in between the low and the high-energy regime. For both of them, we can successfully evaluate the proper limits of SM and use approximated methods to solve the resulting equations, although a satisfactory description in the whole energy range is still missing.

A widely applied strategy to describe the internal structure of hadrons relies on the parton model, which is based on the study of the distribution of partons (i.e. fundamental particles such as quarks and gluon) inside the hadrons. These distributions are extracted from the experiments, by using advanced fitting and modelling methods \cite{AbdulKhalek:2019bux}, and also including spin information within the polarized PDFs. 
However, this methodology is not enough to explain the total spin of the proton. Using up-to-date experimental data, it turns out that only about $30 \%$ of the proton total spin can be explained by the quarks. Moreover, as shown in Refs. \cite{deFlorian:2008mr,deFlorian:2009vb}, data extracted from DIS experiments is not enough to constrain the shape of polarized quark and gluon distributions. A precise description of such distributions is crucial to tackle the \emph{proton spin crisis}, a long-standing problem whose solution is still eluding the efforts of the scientific community.

In order to shed light into possible solutions to the spin crisis and obtain more information about the internal dynamics of hadrons, we need to access to the parton level kinematics in the most clean and unperturbed way. It is a well-known fact that high-energy collisions of hadrons produce a hot and dense medium mainly composed by strongly interacting particles. As a consequence, any particle that couples to QCD partons will suffer from this interaction. The main problem is that such states can not be easily described within the perturbative approach, thus introducing huge uncertainties in the theoretical modelling. Even if there were very recent and precise descriptions of Quark-Gluon Plasma (QGP) evolution by using hydrodynamic and AdS/CFT-inspired models~\cite{Maldacena:1997re,Bertoldi:2007sf}, their consistent combination with the customary perturbative approach is not well understood. For these reasons, one clever alternative to overcome these issues relies in the measurement of final states involving hard photons. These particles are almost transparent to the QGP states, thus allowing to access the parton-level kinematics in a cleaner way.

This work is based on Ref. \cite{deFlorian:2010vy}, and constitutes a step towards a more complete and up-to-date description of the phenomenology of hadron-photon production at colliders. The use of photons as clean probes has been explored in several works to establish patterns of energy loss in heavy ion collisions \cite{Wang:1996yh}, the sensitivity to medium-induced modifications to fragmentation functions (FF) \cite{Arleo:2004xj,Arleo:2006xb,Zhang:2009rn} and constrain photon fragmentation at hadron colliders \cite{Belghobsi:2009hx}, among other studies. Here, we center into the production of a direct photon plus one hadron including next-to-leading order (NLO) corrections due to QCD effects. This is an interesting observable since it involves both PDFs and FFs, which allows to obtain constrains for such distributions. Thus, in Sec. \ref{sec:computation}, we explain the theoretical framework applied to calculate the NLO corrections to the cross-section, as well as the inclusion of sub-dominant QED corrections. Special emphasis is put on the isolation algorithm that allows to efficiently separate the contribution due to direct photon emission, from the one originated from hadron decays. In Sec. \ref{sec:results}, we present some numerical results using up-to-date PDFs and FFs, comparing them to the ones presented in Ref. \cite{deFlorian:2010vy}. Finally, in Sec. \ref{sec:conclusions}, we summarize this presentation and explain future strategies to explore the inner structure of hadrons by using this observable.


\section{Implementation of the computation}
\label{sec:computation}
The computation is based on the parton model for describing hadron-hadron collisions in the high-energy regime. In such kinematic regime, there are factorisation properties~\cite{Collins:1989gx} that allows to apply perturbation theory for computing the cross-section\footnote{The factorisation theorems have been rigorously proven for DIS using the method of operator product expansions (OPE). However, the extension to hadron-hadron collisions is not formally demonstrated, neither the one including also fragmentation into hadrons. In any case, several studies \cite{Catani:2011st,Forshaw:2012bi} have been exploring potential factorisation-breaking issues, and they have shown that these problems might appear beyond NLO when color-charged particles are involved.}. 
Explicitly, the cross-section is described by a convolution between PDFs, FFs and the \emph{partonic} cross-section. All the non-perturbative effects associated to the low-energy interaction inside the hadrons are included within the PDFs and FFs, whilst the partonic cross-section can be computed using the perturbative framework. Thus, in the case of hadron-photon production, we can start writing
\beqn
\nonumber d\sigma_{H_1 \, H_2 \to h \, \gamma}^{\rm DIR} &=& \sum_{a_1 a_2 a_3} \int dx_1 dx_2 dz \, f^{(H_1)}_{a_1}(x_1,\mu_I) f^{(H_2)}_{a_2}(x_2,\mu_I) \, d^{(h)}_{a_3}(z,\mu_F) d\hat\sigma^{\rm DIR}_{a_1\,a_2 \to a_3 \, \gamma}  \, ,
\\ &&
\label{eq:Direct}
\eeqn
with $H_1$ and $H_2$ the hadrons colliding in the initial state, $a_i$ the partons involved in the process and $d\hat\sigma^{\rm DIR}$ the differential partonic cross-section. The function $f^{(H)}_{a}(x,\mu_I)$ represents the PDF associated to the collinear emission of a parton of flavor $a$ from the hadron $H$ with momentum fraction $x$ at the initial factorisation scale $\mu_I$. Analogously, $d^{(h)}_{a}(z,\mu_F)$ represents the density probability function of generating a hadron $h$ with momentum fraction $z$ from the parton $a$, at the final factorisation scale $\mu_F$. Regarding the scale dependence, the partonic cross-section includes terms depending on $\mu_I$, $\mu_F$ and also on the renormalisation scale, $\mu_R$.

This formula assumes that the photon is directly generated in the parton interaction, but additional contributions could arise. For instance, high-energy collisions of hadrons could originate pions, which might eventually decay into photons. Thus, if we are looking for final state high-energy photons, our measurements could also include contributions from this decay process. Due to the quantum nature of the process that we are exploring, it is not possible to identify the true origin of the particles that we look in the detector. Thus, in principle, we must also compute the fragmentation or \emph{resolved} component of the cross-section, i.e.
\beqn
\nonumber d\sigma_{H_1 \, H_2 \to h \, \gamma}^{\rm RES} &=& \sum_{a_1 a_2 a_3 a_4} \int dx_1 dx_2 dz dz' \, f^{(H_1)}_{a_1}(x_1,\mu_I) f^{(H_2)}_{a_2}(x_2,\mu_I) \,
\\ && \times d^{(h)}_{a_3}(z,\mu_F) d^{(\gamma)}_{a_4}(z',\mu_F) d\hat\sigma_{a_1\,a_2 \to a_3 \, a_4}  \, ,
\label{eq:Resolved}
\eeqn
where the parton $a_4$ generates a photon after hadronization. Notice the presence of the parton-to-photon fragmentation function, $d^{(\gamma)}_{a}(z,\mu_F)$. This quantity is not very well constrained experimentally, due to non-perturbative and low-energy effects. Also, strictly speaking, we should include a component originated from a non-perturbative hadronization process leading to the desired final state, i.e. a purely non-perturbative generation of one hadron plus an energetic photon. In any case, beyond the leading order, the separation in direct, resolved or double-resolved is not physical. However, it is possible to efficiently suppress the resolved contribution, relying on the so-called isolation prescriptions. 

\subsection{Isolation and direct photon contribution}
\label{ssec:Isolation}
When QCD parton interact, they usually generates highly-energetic states that very quickly recombines to generate new hadrons. If the detector is hit by a photon, we could use this fact to discriminate its origin. When the photon is generated as a consequence of a QCD-mediated decay, several particles will be produced close to the photon. On the other hand, if the photon was generated in a pure partonic process, then it is expected to leave a clean signal in the detector. It is possible to show that there is a high correlation among direct photon production and \emph{isolated} photons. By definition, isolated photons are those that fulfill certain selection criteria, establishing a separation from this particle to any hadron or jet. Usually, distances are measured within the rapidity-azimuthal plane: if $a=(\eta_1,\phi_1)$ and $b=(\eta_2,\phi_2)$, then
\beq 
\Delta r_{ab} = \sqrt{(\eta_1-\eta_2)^2+(\phi_1-\phi_2)^2} \, ,
\label{eq:Distance}
\eeq
represents the distance between these two points. Different definitions of distance are available, but this one is specially suited for experiments where angular variables can be efficiently measured.

There are several criteria in the market, such as the cone isolation or the smooth isolation prescription. The latest, introduced for the first time in Ref. \cite{Frixione:1998jh}, posses several theoretical advantages. The selection procedure goes as follow:
\begin{enumerate}
    \item Identify each photonic signal in the final state, and draw a cone of radius $r_0$ around it.
    \item If there are not QCD partons inside the cone, the photon is isolated.
    \item If there are QCD partons inside the cone, we calculate their distance to the photon, $r_j$, following Eq. (\ref{eq:Distance}) and then we define the total transverse hadronic energy for a cone of radius $r$ as
    \beq 
    E_T(r) = \sum_{j} E_{T_j} \theta(r-r_j) \, ,
    \label{eq:Energy}
    \eeq
    where $E_{T_j}$ is the transverse energy of the $j$-th QCD parton inside the cone.
    \item Define an arbitrary smooth function $\xi(r)$ which satisfies $\xi(r) \to 0$ for $r\to 0$.
    \item If $E_T(r)<\xi(r)$ for every $r<r_0$ (i.e. for any point inside the fixed cone), then the photon is isolated.
\end{enumerate}
This algorithm is known as \emph{smooth} cone isolation because it forces the QCD partons to have less transverse energy as they are emitted closer to the photons. The fixed cone introduces a static cut-off, and this might lead to problems from the theory side. The smooth cone isolation allows soft gluons in any region of the phase space, leading to a IR-safe definition of the cross-section, specially when higher-order corrections are considered. Also, this prescription completely eliminates the collinear quark radiation, which implies that the fragmentation contribution in Eq. (\ref{eq:Resolved}) can be neglected. In this way, 
\beqn
\nonumber d\sigma_{H_1 \, H_2 \to h \, \gamma} &=& \sum_{a_1 a_2 a_3} \int dx_1 dx_2 dz \, f^{(H_1)}_{a_1}(x_1,\mu_I) f^{(H_2)}_{a_2}(x_2,\mu_I) \, d^{(h)}_{a_3}(z,\mu_F) d\hat\sigma^{\rm ISO}_{a_1\,a_2 \to a_3 \, \gamma}  \, ,
\\ &&
\label{eq:Isolated}
\eeqn
represents the full-contribution to the cross-section when only isolated prompt-photons are measured, removing completely the resolved component and simplifying the calculation.

Finally, we would like to comment on the choice of the isolation prescription. Even if the smooth cone isolation seems to be perfectly suited for theoretical calculations, it has several limitations from the experimental side. Specifically, a high angular resolution is required to implement the condition of Eq. (\ref{eq:Energy}). Thus, most of the experimental collaborations are still relying on fixed cone strategies. However, several studies were performed to compare both approaches, finding that the differences can be minimized or directly neglected for some observables \cite{Cieri:2015wwa,Catani:2018krb}.  

\subsection{Higher-order corrections}
\label{ssec:HO}
Once the observable has been properly defined, we can discuss how to introduce higher-order corrections to Eq. (\ref{eq:Isolated}). Let us consider the Born level kinematic to be given by
\beq
H_1 (K_1) + H_2 (K_2) \to h (K_3) + \gamma (K_4) \, , 
\eeq
with $K_i$ the four-momenta of the different particles in the lab frame. When we apply the parton model, we introduce the momentum fractions $x_1$ and $x_2$ for the incoming partons, and $z$ for the final-state parton that hadronizes into $h$. Thus, we can write
\beq
a_1 (x_1 K_1) + a_2 (x_2 K_2) \to a_3 (K_3/z) + \gamma (K_4) \, , 
\eeq
for the parton kinematics. As usual, the calculation is performed in the partonic center-of-mass frame, and then boosted to the lab frame. 
In this work, we first focus our attention to the QCD corrections to the process $\gamma+h$, up to NLO accuracy. Thus, the partonic cross-section can be expanded according to
\beqn  
\nn d\hat\sigma^{\rm ISO}_{a_1\,a_2 \to a_3 \, \gamma} &=& \frac{\as}{2\pi} \frac{\aem}{2\pi}\, \int d{\rm PS}^{2\to 2} \,  \frac{|{\cal M}^{(0)}|^2(x_1 K_1, x_2 K_2, K_3/z, K_4)}{2 \hat s} \, {\cal S}_2 \, 
\\ \nn &+& \frac{\as^2 }{4\pi^2} \frac{\aem}{2\pi}\, \int d{\rm PS}^{2\to 2} \, \frac{|{\cal M}^{(1)}|^2(x_1 K_1, x_2 K_2, K_3/z, K_4)}{2 \hat s} \,{\cal S}_2 \, 
\\ &+& \frac{\as^2 }{4\pi^2} \frac{\aem}{2\pi} \sum_{a_5} \int d{\rm PS}^{2\to 3} \, \frac{|{\cal M}^{(0)}|^2(x_1 K_1, x_2 K_2, K_3/z, K_4, k_5)}{2 \hat s} \, {\cal S}_3 \,  ,
\label{eq:xsISOLATEDQCD}
\eeqn  
with $\hat s$ the partonic center-of-mass energy, $|{\cal M}^{(0)}|^2$ the squared matrix-element at Born level and $|{\cal M}^{(1)}|^2$ the corresponding one-loop one. ${\cal S}_2$ and ${\cal S}_3$ are the measure functions that implements the experimental cuts and the isolation prescription for the $2\to 2$ and $2 \to 3$ sub-processes, respectively. There are two partonic channels contributing at LO,
\beq
q \bar q \to \gamma g \, , \quad q g \to \gamma q \, ,
\label{eq:PartonicChannelsLO}
\eeq
whilst 
\beq
q \bar q \to \gamma g g \, , \quad q g \to \gamma g q \, , \quad g g  \to \gamma q \bar q \, , \quad q \bar q \to \gamma Q \bar Q \, , \quad q Q \to \gamma q Q \, , 
\label{eq:PartonicChannelsNLO}
\eeq
are all the QCD channels contributing at NLO. The opening of new channels could result in enhanced effects due to gluon PDFs.

In order to implement the NLO QCD corrections, we relied on the FKS subtraction algorithm \cite{Frixione:1995ms}, that splits the real emission phase-space into different regions containing non-overlapping singularities, and cancels them with local IR counter-terms. The renormalisation was performed within the $\overline{\rm MS}$ scheme. 

On top of that, we also explored the impact of QED corrections. Going back to Eq. (\ref{eq:xsISOLATEDQCD}), we appreciate that the NLO QCD contributions and LO QED corrections might have a similar weight, due to the relation $\as^2 \, \aem \approx \aem^2 \approx {\cal O}(10^{-4})$. Of course, photon PDF is suppressed w.r.t. the QCD ones because of the sub-leading role of the electromagnetic interactions in the determination of the inner structure of hadrons. In any case, it constitutes a non-negligible correction due to the high-accuracy in nowadays experiments. Thus, we add to Eq. (\ref{eq:xsISOLATEDQCD}) \beqn  
\nn d\hat\sigma^{\rm ISO, QED}_{a_1\,a_2 \to a_3 \, \gamma} &=& \frac{\aem^2}{4\pi^2}\, \int d{\rm PS}^{2\to 2} \,  \frac{|{\cal M}^{(0)}_{QED}|^2(x_1 K_1, x_2 K_2, K_3/z, K_4)}{2 \hat s} \, {\cal S}_2 \,  ,
\label{eq:xsISOLATEDQED}
\eeqn 
associated to the new partonic channel
\beq
q \gamma \to \gamma q \, , \quad q \bar q \to \gamma \gamma  \, .
\label{eq:PartonicChannelsLOQED}
\eeq
Regarding the second channel, we will neglect its contribution by looking into charged hadrons in the final state. This is because such sub-process involves the fragmentation of a photon into a hadron, which is also suppressed by the fact that electromagnetic interactions are sub-leading in the hadronization process. Even more, we assume that electromagnetic contribution to the hadronization is almost negligible compared to photon-production from hadrons\footnote{We defer for future works a deeper study of this assertion.}.


\section{Numerical simulations}
\label{sec:results}
The computation described in Sec. \ref{sec:computation} was implemented in a Monte Carlo integrator, based on the one developed in Ref. \cite{deFlorian:2010vy}. For the isolation algorithm, we used the function
\beq 
\xi(r) = \epsilon_\gamma E_T^{\gamma} \, \left(\frac{1-\cos(r)}{1-\cos{r_0}}\right)^4 \, ,
\eeq 
with the parameters $\epsilon_\gamma=1$, $r_0=0.4$ and the photon transverse energy $E_T^{\gamma}$. The average of the photon and hadron transverse energy was used as the typical energy scale of the process, i.e.
\beq  
\mu \equiv \frac{p_T^{h}+p_T^{\gamma}}{2} \, ,
\eeq
and we set by default $\mu_I=\mu_F=\mu_R\equiv \mu$.

Due to the fact that the production probability of heavy hadrons is very suppressed, we will focus on the the process $p + p \to \pi + \gamma$, including the possibility of charge selection for the pion. Regarding the cuts, our default configuration corresponds to the ones used by the PHENIX detector: 
\begin{itemize}
  \item Pion and photon rapidities are restricted to $|\eta|\leq 0.35$.
  \item The photon transverse momentum fulfills $5 \, {\rm GeV} \leq p_T^{\gamma} \leq 15 \, {\rm GeV}$.
  \item Pion transverse momentum must be larger than 2 GeV.
  \item We consider full azimuthal coverage, i.e. no restriction on $\{\phi^\pi, \phi^\gamma\}$, as a simplification of the real detectors.
\end{itemize}
The lower cut on the pion transverse momentum is set to reduce the contamination from the non-perturbative processes. A similar argument holds for the limitation in the photon energy range. Regarding the center-of-mass energy of the hadron collisions, we use by default $E_{CM} = 200 \, {\rm GeV}$, although we also explored the TeV region accessible by LHC, setting $E_{CM} = 13 \, {\rm TeV}$.

As already proposed in Ref. \cite{deFlorian:2010vy}, we restrict to $\Delta \phi=|\phi^\pi - \phi^\gamma|\geq 2$ in order to keep those events were the photon and the pion are produced in an almost back-to-back configuration. This is because the LO kinematics, in the partonic center-of-mass frame, allows only back-to-back events; thus, imposing that restriction on $\Delta \phi$ is equivalent to look around the Born kinematics.

Finally, we performed an update in the PDF and FF available in the original version of the code \cite{deFlorian:2010vy}. In particular, we switched from the stand-alone implementations of the different PDFs to the \texttt{LHAPDF} framework \cite{Buckley:2014ana}. This allows to unify the treatment of the PDFs, and simplify the numerical evaluations in the different scenarios. Besides, we implemented in the code the updated set of FFs, \texttt{DSS2014} \cite{deFlorian:2014xna}.

\subsection{Phenomenology and results}
\label{ssec:phenomenology}
\begin{figure}[h!]
    \centering
    \includegraphics[width=0.75\textwidth]{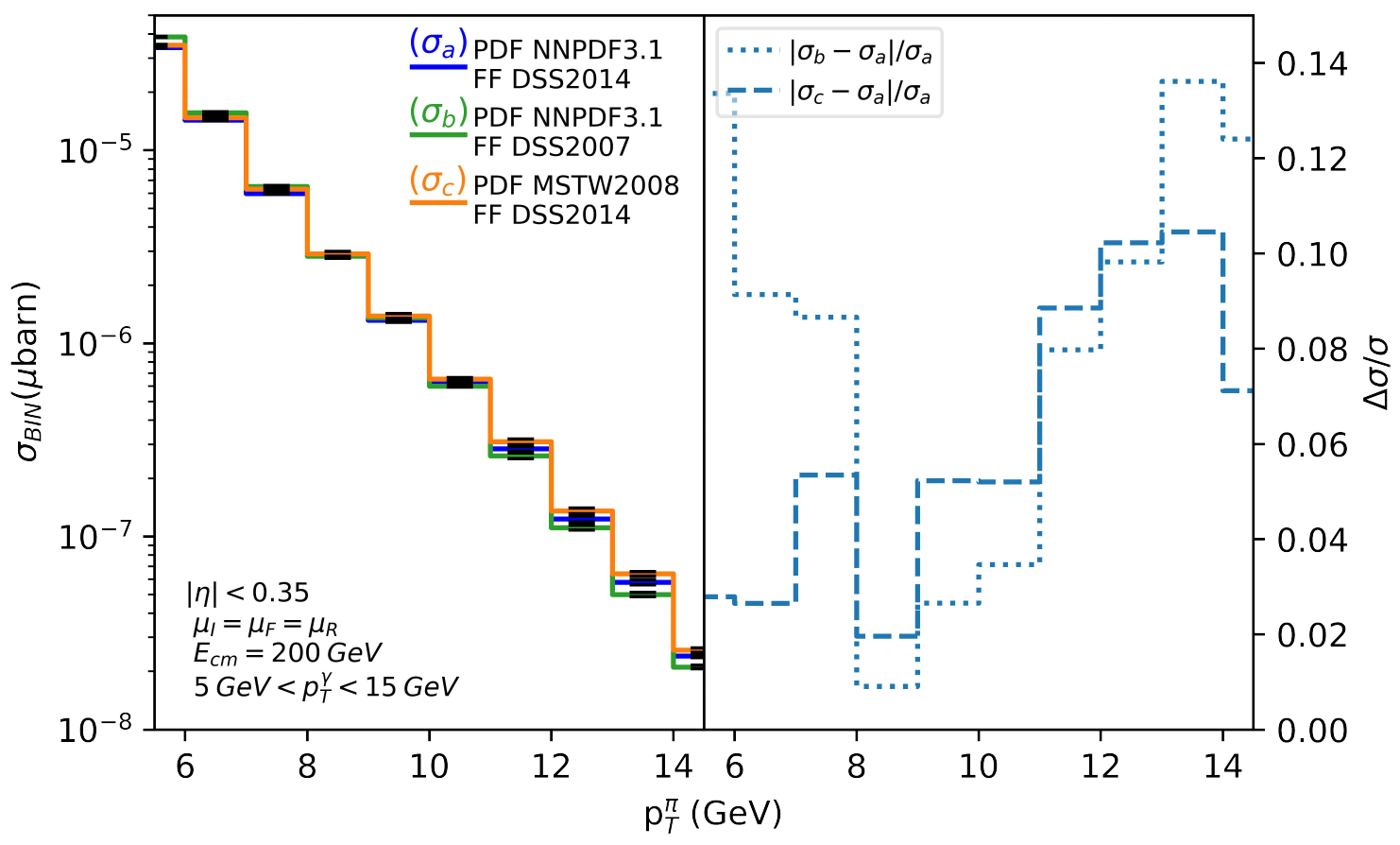}
    \caption{NLO QCD corrections to the $p_T^{\pi}$ distribution of $pp\to \gamma + \pi^+$, for PHENIX kinematics ($E_{CM} = 200 \, {\rm GeV}$) in three different scenarios. In the right panel, we show the relative difference w.r.t. the default configuration (i.e. \texttt{NNPDF3.1} PDF and \texttt{DSS2014} FF).}
    \label{fig:Figura1}
\end{figure}

\begin{figure}[h!]
    \centering
    \includegraphics[width=0.75\textwidth]{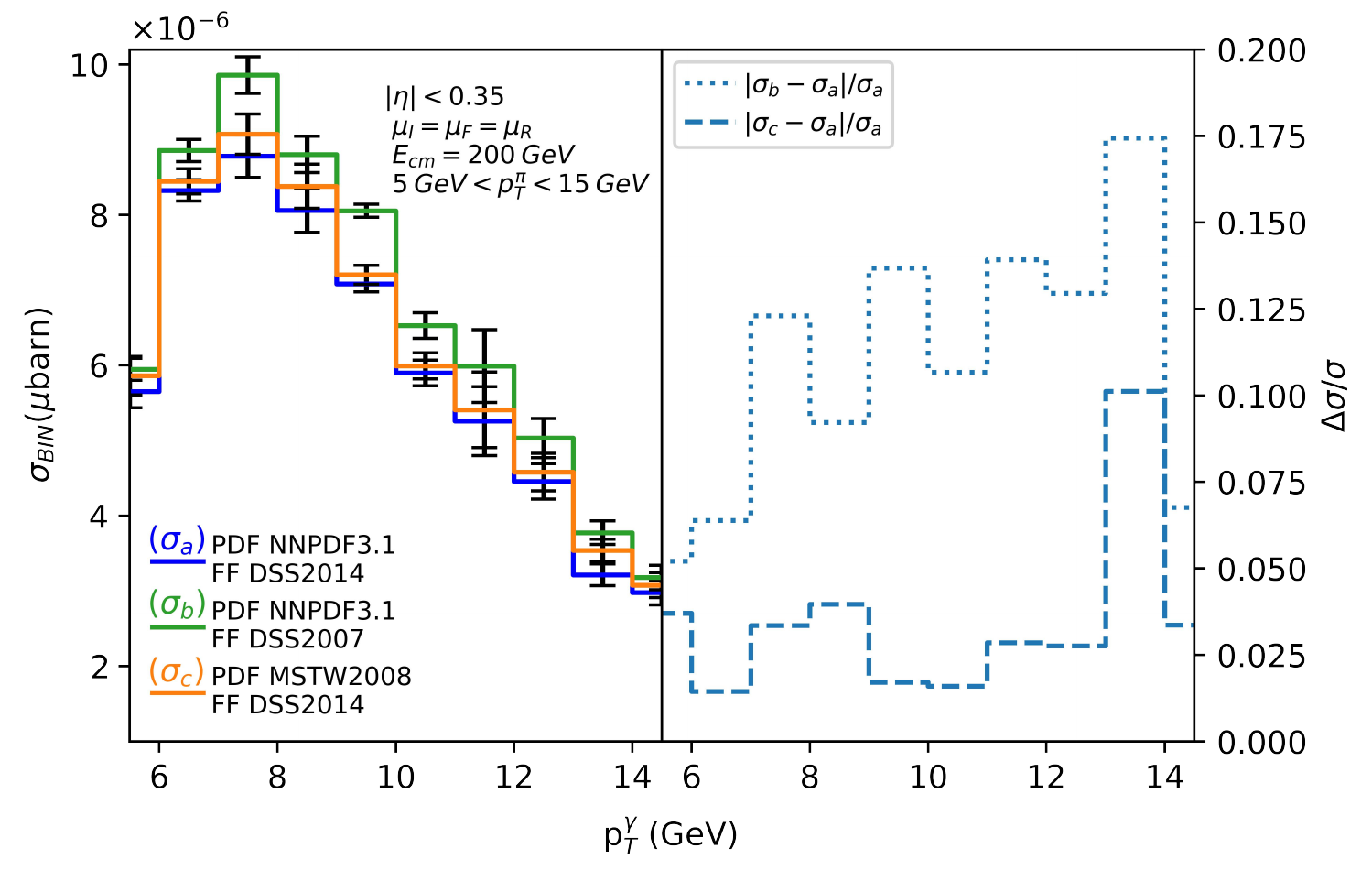}
    \caption{NLO QCD corrections to the $p_T^{\gamma}$ distribution of $pp\to \gamma + \pi^+$, for PHENIX kinematics ($E_{CM} = 200 \, {\rm GeV}$) in three different scenarios. In the right panel, we show the relative difference w.r.t. the default configuration (i.e. \texttt{NNPDF3.1} PDF and \texttt{DSS2014} FF).}
    \label{fig:Figura2}
\end{figure}

In first place, we reproduced the old results obtained with \texttt{MSTW2008NLO} PDFs \cite{Martin:2009iq} and \texttt{DSS2007} fragmentations \cite{deFlorian:2007ekg}, but using the new Monte Carlo implementation within the \texttt{LHAPDF} framework. Then, we explored the effects introduced by switching to novel versions of the PDFs and FFs. In particular, we considered three configurations:
\begin{enumerate}
    \item $\sigma_a$: \texttt{NNPDF3.1} and \texttt{DSS2014} (default up-to-date simulation)
    \item $\sigma_b$: \texttt{NNPDF3.1} and \texttt{DSS2007} (effects in the hadronization)
    \item $\sigma_c$: \texttt{MSTW2008} and \texttt{DSS2014} (effects in the parton distributions)
\end{enumerate}
using always the corresponding sets with NLO QCD corrections. In first place, we considered the kinematics for the PHENIX experiment, focusing on $pp$ collisions at $E_{CM}=200 \, {\rm GeV}$. In Figs. \ref{fig:Figura1} and \ref{fig:Figura2}, we plot the differential cross-section as a function of $p_T^{\pi}$ and $p_T^{\gamma}$, respectively. We restricted our attention to the NLO QCD corrections for the process $pp\to \gamma+\pi^+$ and used the default scale choice (as explained in the previous section). 

Regarding the $p_T^{\pi}$ spectrum for the three scenarios, we found that they are pretty similar. In fact, the absolute difference is smaller than $15 \, \%$ in the range $5 \, {\rm GeV} \leq p_{T}^{\pi} \leq 15 \, {\rm GeV}$. From the right plot in Fig. \ref{fig:Figura1}, we notice that the \texttt{MSTW2008} PDF tends to slightly enhance the high-$p_T$ region (i.e. $p_T^{\pi}\approx 13 \, {\rm GeV}$), whilst the effect of the \texttt{DSS2007} FF goes in the opposite direction. Namely, \texttt{DSS2007} gives a larger cross-section for low-$p_T$ and a smaller for high-$p_T$, reaching a relative difference of ${\cal O}(10 \, \%)$. In any case, it seems that the integrated effect seems to compensate across the whole range of $p_T^{\pi}$.

On the other hand, if we look at the $p_T^{\gamma}$ spectrum, we found greater deviations from the default configuration. As shown in the right plot of Fig. \ref{fig:Figura2}, the cross-section is about $12 \, \%$ bigger when using the \texttt{DSS2007} fragmentations, and the trend is increasing with the photon transverse momentum. When we switch to \texttt{MSTW2008} PDFs, we also find that the cross-section is higher than for the default configuration, and the trend is increasing with $p_T^\gamma$: the relative difference is ${\cal O}(7 \, \%)$.

\begin{figure}[h!]
    \centering
    \includegraphics[width=0.75\textwidth]{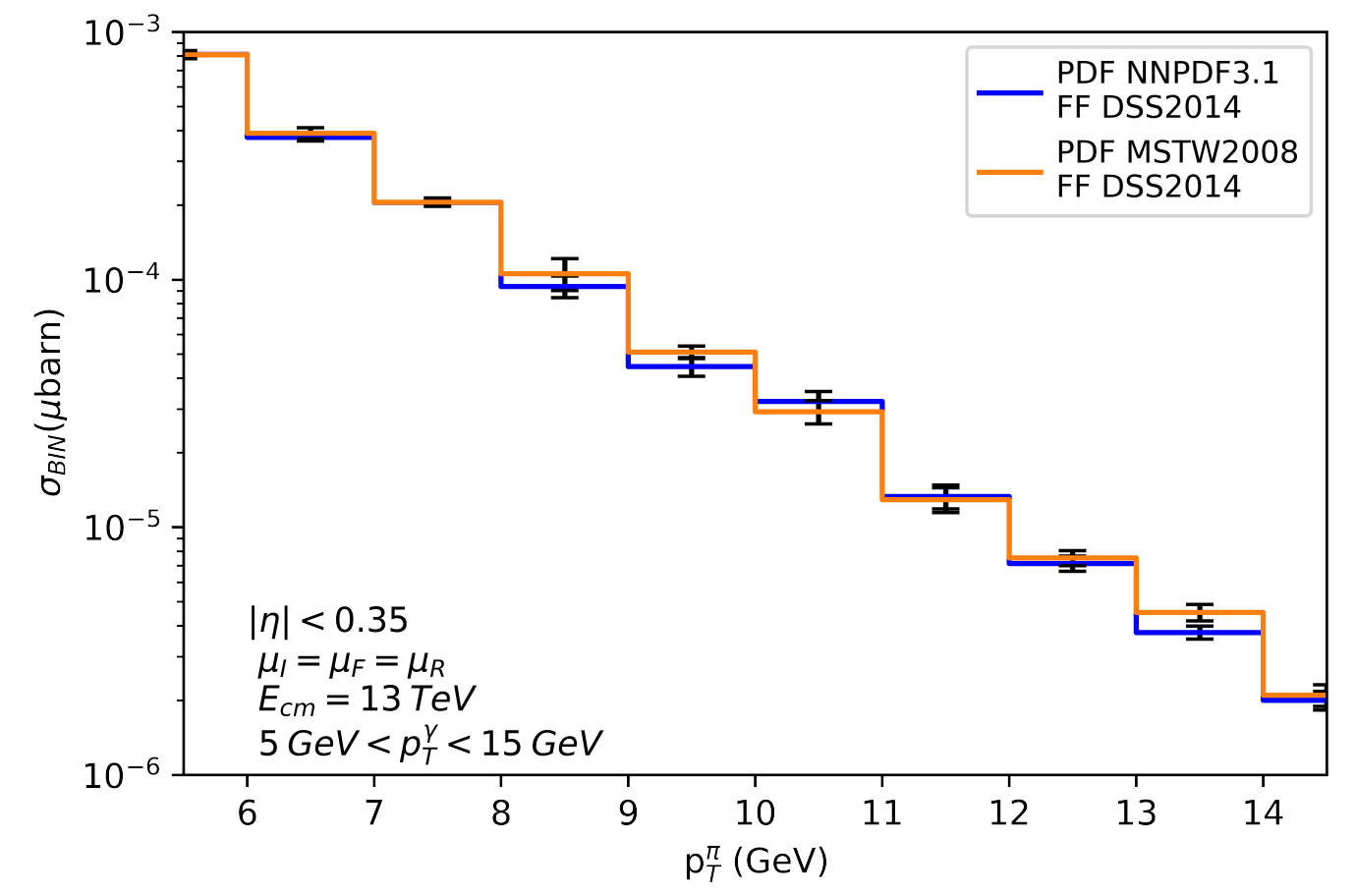}
    \caption{NLO QCD corrections to the $p_T^{\pi}$ distribution of $pp\to \gamma + \pi^+$, for LHC kinematics ($E_{CM} = 13 \, {\rm TeV}$, angular constraints identical to those used for PHENIX) in three different scenarios. The relative differences are shown in the right panel.}
    \label{fig:Figura3}
\end{figure}

The comparison among the three PDF-FF scenarios show that there is room to impose tighter constraints on both the PDFs and FFs from PHENIX kinematic. However, given the fact that new colliders are extending the available energy range, we studied the differences in the $p_T^{\pi}$ spectrum at LHC, namely setting $E_{CM} = 13 \, {\rm TeV}$ whilst keeping the same angular resolution used for the PHENIX experiment. The results are shown in Fig. \ref{fig:Figura3}, where we restricted the attention to $\sigma_a$ and $\sigma_c$. Namely, we compared \texttt{NNPDF3.1} (blue line) and \texttt{MSTW2008} (orange line), keeping the same fragmentation functions. The differences are around $10 \, \%$, with an enhancement of the cross-section for $p_T^{\pi} \approx 13 \, {\rm GeV}$ for $\sigma_a$ w.r.t. $\sigma_c$. As already observed in Figs. \ref{fig:Figura1} and \ref{fig:Figura2}, . It is worth noticing that we did not use \texttt{DSS2007} for LHC energies because most of the events involve momentum fractions laying outside the validity range of the interpolator.

\begin{figure}[h!]
    \centering
    \includegraphics[width=0.47\textwidth]{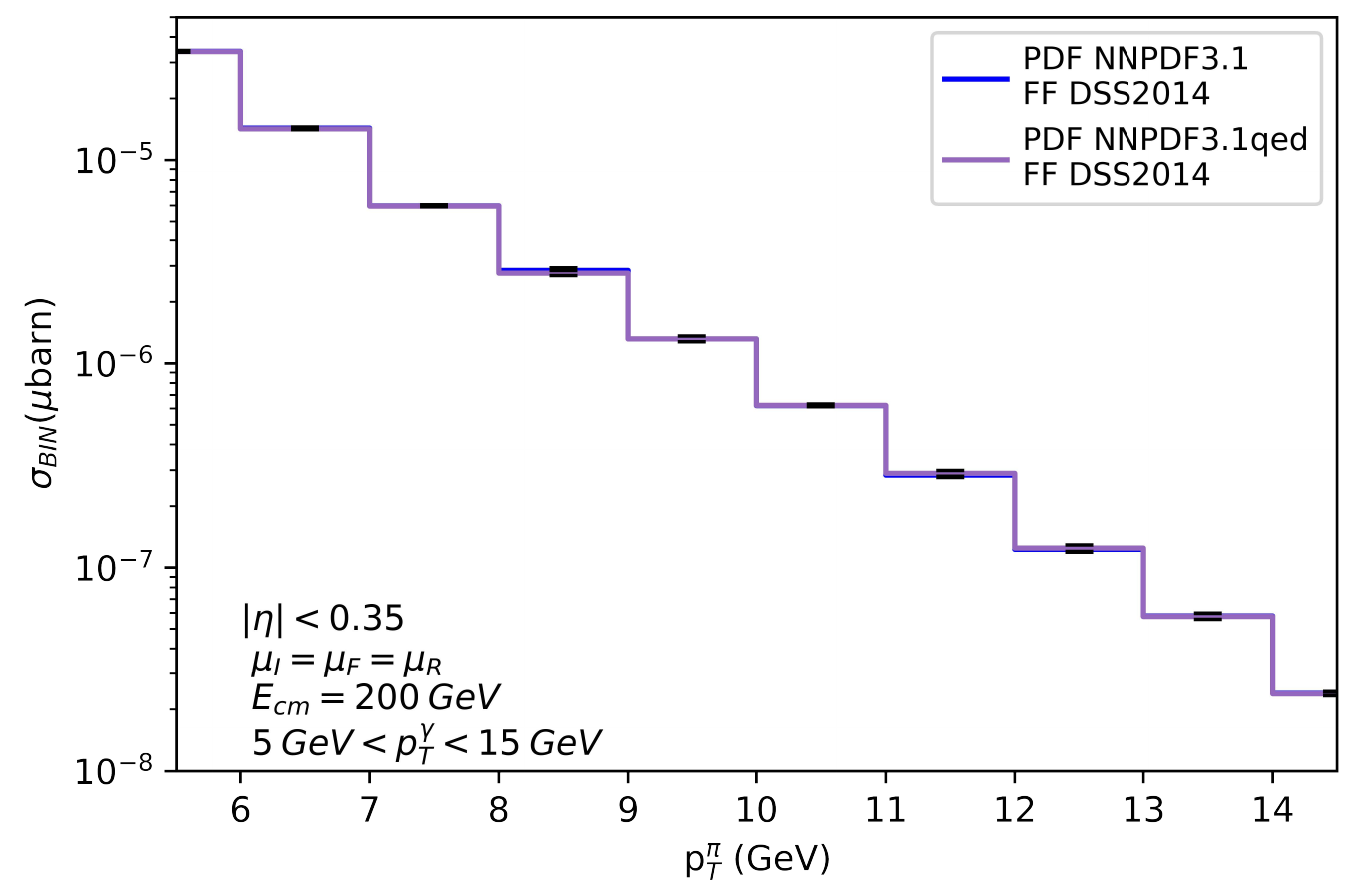} \ \includegraphics[width=0.47\textwidth]{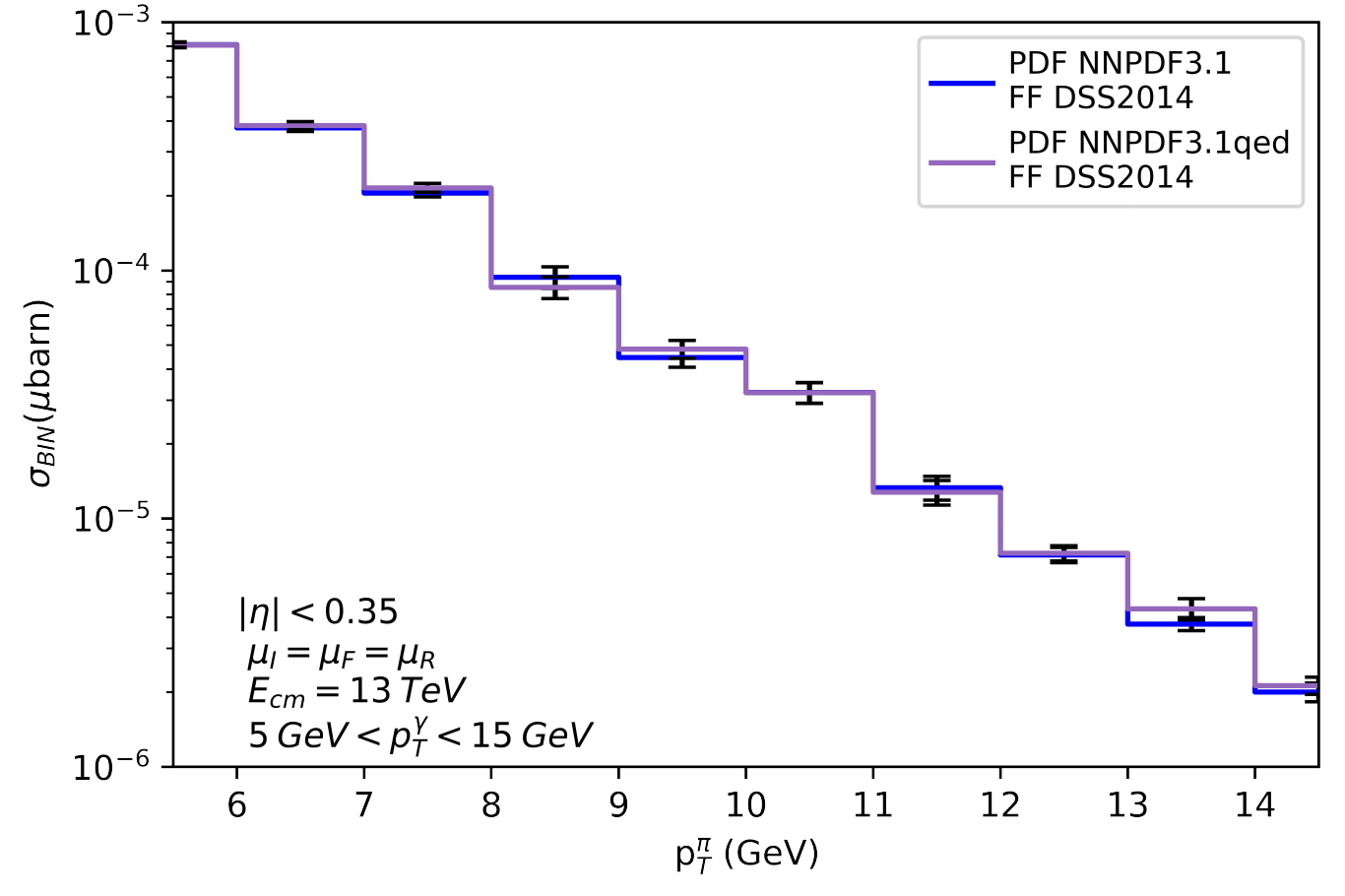}
    \caption{Comparison of the combined NLO QCD + LO QED (blue line) vs the pure NLO QCD corrections to the $p_T^{\pi}$ distribution of $pp\to \gamma + \pi^+$. In the left, we consider PHENIX kinematics ($E_{CM}=200 \, {\rm GeV}$), whilst we switched to LHC energies ($E_{CM} = 13 \, {\rm TeV}$) in the right plot. We use as a shorthand notation \texttt{NNPDF3.1qed} for \texttt{NNPDF3.1luxQEDNLO}.}
    \label{fig:Figura4}
\end{figure}

Finally, we explored the impact of the combined NLO QCD + LO QED corrections to the process $pp \to \gamma + \pi^+$. In order to deal with the photon distribution, we switched to \texttt{NNPDF3.1luxQEDNLO} \cite{Manohar:2017eqh,Bertone:2017bme,Buonocore:2020nai}, which also include lepton densities\footnote{In our analysis, we are leaving aside lepton initiated processes, since their contribution is very suppressed compared to the QCD one. The inclusion of such corrections is deferred for future studies.}. In Fig. \ref{fig:Figura4}, we compare the $p_T^{\pi}$ spectrum at $E_{CM}=200 \, {\rm GeV}$ (left) and $E_{CM}=13 \, {\rm TeV}$ (right), for proton-proton collisions extrapolating the typical PHENIX cuts. We found corrections of ${\cal O}(2 \, \%)$ for $E_{CM}=200 \, {\rm GeV}$, whilst they increase to ${\cal O}(9 \, \%)$ for LHC energies. This is partially due to the behaviour of the QED and QCD couplings, which goes in opposite directions as the process energy increases. The weight of QED corrections tends to be slightly higher in the high $p_T$-region, although it seems to be more affected by the center-of-mass energy of the collision. 

\section{Conclusions and outlook}
\label{sec:conclusions}
In this article, we discussed the phenomenology of photon-hadron production at hadron colliders, centering in the process $pp \to \gamma + \pi^+$. We included NLO QCD and LO QED corrections to keep under control effects of order $\aem^2 \approx \as^2 \aem$. The results were compared with previous studies done in Ref. \cite{deFlorian:2010vy}.

As a first step, we updated the Monte Carlo implementation to include the unified framework \texttt{LHAPDF} and the new set of pion FF of \texttt{DSS2014}. We carefully studied the dependence of the PDF and FF sets, by looking into the $p_T^{\pi}$ and $p_T^{\gamma}$ distributions. We found reasonable deviations (i.e. ${\cal O}(10 \, \%)$ on average), although our preliminary studies suggest a stronger sensibility in the $p_T^{\gamma}$ distribution.

Then, we included the LO QED corrections by opening the partonic channel $q\gamma$. We used the \texttt{NNPDF3.1luxQEDNLO} PDF set, and tested the impact of these contributions to the $p_T^{\pi}$ spectrum. We restricted our attention to the production of positive pions, since we claim that such choice would heavily suppress any contribution associated to the hadronization process $\gamma \to \pi$. Comparing the corrections at $pp$ colliders, we found small but still non-negligible corrections: ${\cal O}(2\, \%)$ for PHENIX and ${\cal O}(8\, \%)$ for LHC center-of-mass energies.

The results presented in this article suggest that hadron+photon production might be a useful process to impose tighter constraints on both PDFs and FFs. An extensive study for different observables could lead to an enhanced sensitivity on the PDFs, through a careful reconstruction of the partonic momentum fractions. Further developments on this directions are being done by our group.

\section*{Acknowledgements}
This research was supported in part by COST Action CA16201 (PARTICLEFACE).
The work of D. F. R.-E. and R.J. H.-P. is supported by CONACyT through the Project No. A1- S-33202 (Ciencia Basica). Besides, R.J. H.-P. is also funded by Ciencia de Frontera 2021-2042 and Sistema Nacional de Investigadores from CONACyT.

\bibliographystyle{JHEP}

\providecommand{\href}[2]{#2}\begingroup\raggedright\endgroup

\end{document}